\documentclass[conference]{IEEEtran}

\IEEEoverridecommandlockouts
\usepackage[utf8]{inputenc}
\usepackage{cite}
\usepackage{amsmath,amssymb,amsfonts}
\usepackage{algorithmic}
\usepackage{framed}
\usepackage{graphicx}
\usepackage{wrapfig}
\usepackage{textcomp}
\usepackage{xcolor}
 \usepackage[colorlinks=true,
        linkcolor=black,
        citecolor=black,
        filecolor=black,
        urlcolor=black,
        bookmarks=true,
        bookmarksopen=true,
        bookmarksopenlevel=3,
        plainpages=false,
        pdfpagelabels=true]{hyperref}
\def\BibTeX{{\rm B\kern-.05em{\sc i\kern-.025em b}\kern-.08em
    T\kern-.1667em\lower.7ex\hbox{E}\kern-.125emX}}

\usepackage{multirow}

\usepackage{array}
\usepackage{colortbl}
    
\usepackage{ifthen}

\newboolean{auxinfo}
\setboolean{auxinfo}{true}   

\newboolean{showoutline}
\setboolean{showoutline}{true}

\newboolean{showcomment}

\setboolean{showcomment}{true} 

\ifthenelse{\boolean{auxinfo}}
  {}
  {
    \setboolean{showoutline}{false}
    \setboolean{showcomment}{false}
  }

\usepackage{xcolor}
\usepackage{amssymb}

\ifthenelse{\boolean{showcomment}}
   {\newcommand{\nb}[2]{\fcolorbox{gray}{yellow}{\bfseries\sffamily\scriptsize#1}{\sf\small$\blacktriangleright${\em #2}$\blacktriangleleft$}}
   \newcommand{\working}[1]{\fcolorbox{gray}{yellow}{{\bf #1}\emph{\scriptsize---in progress---}}}
   \newcommand{\TBD}[1]{\fcolorbox{gray}{yellow}{{\bf #1}\textbf{TBD}}} 
  }
  {\newcommand{\nb}[2]{}{}
   \newcommand{\working}[1]{}
   \newcommand{\TBD}[1]{} 
  }

\usepackage{todonotes}
\ifthenelse{\boolean{showoutline}}{
	\newcommand{\outline}[3]{
		~\newline 
		\fcolorbox{red}{white}{
			\parbox{\columnwidth}{
				\ifthenelse{\equal{#1}{}}{
					\ifthenelse{\equal{#2}{}}{
						\noindent\colorbox[rgb]{0.65,0.16,0}{\textcolor[rgb]{1,1,1}{\textbf{Outline}}}
					}{
						\colorbox[rgb]{0.65,0.16,0}{\textcolor[rgb]{1,1,1}{\textbf{Outline -- Responsible: #2}}}
					}
				}{
					\ifthenelse{\equal{#2}{}}{
						\noindent\colorbox[rgb]{0.65,0.16,0}{\textcolor[rgb]{1,1,1}{\textbf{#1 page(s)}}}
					}{
						\colorbox[rgb]{0.65,0.16,0}{\textcolor[rgb]{1,1,1}{\textbf{#1 page(s) -- Responsible: #2}}}
					}
				}
				#3
			}
		}
	}
}{
	\newcommand{\outline}[3]{}
}

\newcommand\defauxcomm[1]{
       \expandafter\newcommand\csname #1\endcsname[1]{\nb{#1}{##1}}
       \expandafter\newcommand\csname WK#1\endcsname{\working{#1}}
       \expandafter\newcommand\csname TBD#1\endcsname{\nb{#1}}
    } 
   
\defauxcomm{KS}
\defauxcomm{MN}

\usepackage[normalem]{ulem}
\ifthenelse{\boolean{showcomment}}
    {\newcommand{\strike}[1]{\textcolor{red}{\sout{#1}}}}
    {\newcommand{\strike}[1]{}}

\definecolor{tablehead}{HTML}{F9F5F9} 

\begin{document}

\title{Modelling Interrelations Between Agile Practices: The Agile Map} 



\author{
\IEEEauthorblockN{Thomas Hansper, Kevin Phong Pham}
\IEEEauthorblockA{University of Applied Sciences FHDW Hannover \\
Hannover, Germany \\
\{firstname.lastname\}@edu.fhdw.de}
\and
\IEEEauthorblockN{Michael Neumann}
\IEEEauthorblockA{University of Applied Sciences Hannover \\
\textit{Dpt. of Business Information Systems}\\
Hannover, Germany \\
michael.neumann@hs-hannover.de}
}


\maketitle

\begin{abstract}
\textit{Context:} Agile methods are defined through guidelines comprising various practices intended to enable agile ways of working. These guidelines further comprise a specific set of agile practices aiming to enable teams for an agile way of working. However, due to its wide-spread use in practice we know that agile practices are adopted and tailored intensively, which lead to a high variety of agile practices in terms of their level of detail. 
\textit{Problem:} A high variety of agile practices can be challenging as we do not know how different agile practices are interrelated with each other. To be more precise, tailoring and adopting agile practices may lead to the challenge, that the combinatorial use of several agile practices can only be successful to a limited extent, as practices support or even require each other for a effective use in practice.
\textit{Objective:} Our study aims to provide an enabler for this problem. We want to identify interrelations between agile practices and describe them in a systematic manner. \textit{Contribution:} The core contribution of this paper is the Agile Map, a theoretical model describing relations between agile practices following a systematic approach aiming to provide an overview of coherences between agile practices. The model aims to support practitioners in selecting and combining agile practices in a meaningful way.
\end{abstract}


\section{Introduction}
\label{Sec1:Intro}
Agile methods, such as Scrum~\cite{Schwaber.2020}, are widely adopted in software development~\cite{VersionOne.2023}. Organizations worldwide implement agile approaches to respond to the increasing dynamics of markets and address challenges such as rapidly evolving requirements. Agile methods are value-driven frameworks that emphasize team collaboration and interaction~\cite{Beck.2001}. Typically, guidelines define how to apply and use agile methods. For instance, the Scrum Guide~\cite{Schwaber.2020} define a structured framework comprising artifacts, roles, agile practices, and micro-techniques. This guideline-based approach to software process definition is also characteristic of other agile methods, such as eXreme Programming (XP)~\cite{Beck.2000}.

However, due to the increased use of agile methods in practice and in various different contexts like industries, company sizes, or organizational cultures, we observe an increased adaption of agile approaches and their elements~\cite{Kuhrmann.2022,Sidky.2007}. Several studies deal with the tailoring of agile practices and provide evidence on the high variety of various different agile elements used in practice (e.g., \cite{Diebold.2014,Neumann.2022}). This high variety leads to specific challenges and problems like a lack of a systematic and common understanding of the terminology related to agile methods~\cite{Neumann.2021a}. 

In the last decade, several studies (e.g., \cite{Diebold.2015a,Neumann.2023,Neumann.2025}) deal with the influences on agile practices or the impact of agile practices on the outcome of software projects. Though these models and their underlying theories are important for the research and practitioner communities in the area of  agile software development, these models lack on an important aspect: The question how a sensefull combination of adopting agile practices should be considered. As described above we know, that a specific set of agile practices and rules to apply them in practice is defined in the guidelines of agile methods~\cite{Schwaber.2020}. Thus, one may assume that agile practices may affect each other when applying them and we searched for studies dealing with that topic. However, tailoring agile methods and their elements for their use in practice is the norm~\cite{Neumann.2022}. And thus, organizations are constantly challenged by selecting the best suited agile practices for their needs or even adapting them~\cite{Diebold.2014}. For a successfull use of agile methods in practice it is of high importance to be aware of the interrelationships of the underlying roles, practices, and micro-techniques~\cite{Neumann.2021a}. For example, what are the effects applying a Daily Stand-Up Meeting without an iteration-based approach or a Kanban Board. 

The above motivates the objective of our study, which is refined by the following research questions: 
\begin{itemize}
    \item \textbf{RQ:} How can we describe the interrelationships between agile practices in a systematic manner?
\end{itemize}

This paper is structured as follows: In Section~\ref{Sec2Background}, we provide a background on agile practices, followed by an overview of the related work in Section~\ref{Sec3:RelWork}. Section~\ref{Sec4:RelationsbetweenAP} presents our analysis approach and how we created the agile practices relations repository. Before we close the paper with a summary and future work outlook in Section~\ref{Sec6:Conclusion}, we present the core contribution of this paper, the Agile Map a model describing the coherences between agile practices in a systematic manner.

\section{Theoretical Background}
\label{Sec2Background}

\subsection{Agile Methods \& Practices}
A software development process is a systematic description how we create a result using a specific process model by involving specific roles applying specific activities by the support of tools and methods~\cite{SweBok.2014}. Agile practices, which may understood as activities, are a core component in agile methods and of high importance for their successful application in practice~\cite{Schon.2017}. They describe on a detailed level how an agile approach can (or even should) be applied to shape a team for providing transparency of their progress~\cite{Diebold.2015}. Several agile practices enable teams for a continuous improvement process. Thus, these practices play a crucial role for software process improvement~\cite{Kurapati.2012}. For the well-known agile method Scrum~\cite{Schwaber.2020}, the agile practices are used for providing transparency of the progress and process and thus, the opportunity to inspect and adapt the process in use (by adapting a retrospective) or the ongoing output/outcome progress (by using daily or review meeting). However, several other agile practices exist focusing on planning the development tasks, refining requirements for the upcoming iterations, or even describing the art of collaborative software development activities by collective code ownership (e.g., defined for XP by Beck~\cite{Beck.2000}). We can conclude that agile practices are a central element in agile methods, but differ related to the context and in their level of detail.

\subsection{Challenges \& Problems}
Beginning with challenges related to the high variety of agile practices in literature and practice, we discuss the challenges occured by a non-common understanding. In the past decade, several authors investigated this phenomenon aiming to provide a solution for upcoming challenges by non-aligned terminology in the field of agile software development (e.g., \cite{Kurapati.2012,Diebold.2014, Neumann.2021a}). While screening the guidelines of agile methods and studies dealing with agile practices, it often remains unclear what an agile practice exactly is and what not. For instance, some authors understand agile practices as singular activities in software development processes (e.g., \cite{Abrahamsson.2002,Diebold.2016,Sidky.2007}) while others understand them as similar to agile methods. Furthermore, we know that agile practices are named differently with regard to specific agile methods like Scrum. A good example for this situation is the \textit{Daily Meeting}, which may also named as\textit{ Daily Stand-Up meeting} (e.g., used in Kanban~\cite[p.~90]{Anderson.2011}) or in Scrum \textit{Daily Scrum}~\cite{Schwaber.2020}. Thus, there are agile practices with different names in the literature that have the same purpose in practice.Consequently, we are challenged by potential misinterpretions as no common terminology exists. 

\begin{framed}
\textbf{Definition \textit{Agile Practice~\cite{Neumann.2021a}}}: Agile practices are carrying out activities using tools and methods. The implementation can differ depending on the context of application.
\end{framed}

To stay in line with existing literature, we decided to use the \textit{Tree of Agile Elements} presented by Neumann~\cite{Neumann.2021a} as the reference model provides a common understanding of important terms related to agile practices for this study. According to the author, agile elements are defined as roles, artifacts, or activities. For our study, the dimension of agile activities are of importance as it describes agile practices on an abstract level. We follow the definition of agile practices given in~\cite{Neumann.2021a}.

The second challenge we want to focus on is related to the high variety of existing agile practices reported in literature and used in practice, e.g., due to various the levels of details exist. Agile practices may be described on a very specific level how to apply them. A \textit{Daily Meeting} or estimation practices like \textit{Planning Poker} are good examples for very specific defined  agile practices. In contrast, we know that other practices are defined on a more abstract level like retrospective or even review meetings. This situation often leads to a multitude of special forms of the very abstract agile practice, which are not clearly defined and therefore difficult to differentiate from one another. For example, the Agile Alliance~\cite{AgileAlliance.2021} defines a glossar of agile practices consisting of 75 different agile practices from different sources as agile methods like scrum or context related areas like product management. The high variety of agile practices has been studied in recent years e.g., by trying to identify an overview of current used agile practices (~\cite{Diebold.2014,Jalali.2012,Kurapati.2012,Neumann.2022} to name a few). However, agile practices evolve over time as it is totally common to tailor them for their use in practice with regard to the underlying context (e.g., an organizational setting or cultural aspects)~\cite{Wang.2012b}. Also agile practices are often used in combination by applying micro-techniques (i.e., retrospectives). 

\section{Related Work}
\label{Sec3:RelWork}
We searched for both primary and secondary studies dealing with topics closely relate to our topic. In summary, we found that research focusing primarly on relations between agile practices and the underlying effects is lacking. Thus, we decided to expand our literature search to research results dealing with systematic descriptions of effects on agile practices or from agile practices on further aspects like project outcomes. 

The first model we identified is the Model of Cultural Impact on Agile Methods (MoCA)~\cite{Neumann.2022}. The MoCA model was developed to systematically describe the cultural influences on agile practices. Its scientific foundation is based on two dimensions: cultural characteristics and agile elements, as well as the defined interactions between them. The cultural dimension draws on widely used cultural models in software engineering. The agile elements dimension was derived from the results of a tertiary study aimed at providing an up-to-date and comprehensive list of agile practices~\cite{Neumann.2022}. The influence between the two dimensions is characterized as either positive or negative, depending on how the agile element is applied in relation to the guidelines of the respective agile method.

Also, we found the Agile Practices Impact Model (APIM) describing the effects of agile practices on projekt characteristics and outcomes in a systematic manner~\cite{Diebold.2015a}. APIM serves as a foundational framework for assessing agile capability. It links agile practices to distinct impact characteristics, which are framed as process improvement objectives. Although the model incorporates the influence of specific practices, its primary emphasis is on evaluating outcomes. Conceptually, APIM is grounded in three core components: agile practices, impact characteristics, and the relationships between them, which are further detailed through the use of influence factors. These relationships are expressed in binary terms, representing either a positive or negative effect.

Additionally to these findings, we identified research dealing with the high diversity of agile practices in use~\cite{Diebold.2014,Neumann.2022}. Several papers providing overviews of agile practices used in practical environments and contexts (e.g., \cite{Diebold.2014,Kurapati.2012}). Furtheremore, we found approaches to synthesize agile practices on an abstract level aiming to give an up-to-date overview of the applied agile practices by organizations~\cite{Neumann.2022}. Nevertheless, all thess identified studies focus not on the challenge to better understand how specific agile practices influence each other when they are applied in practice. 

Finally and to the best of our knowledge, we could not identify a primary or secondary study dealing with coherences between agile practices providing an in-depth understanding of these interrelationships. 

\section{Relations between Agile Practices}
\label{Sec4:RelationsbetweenAP}
To be able to analyze and define relationships between agile practices, we need an up-to-date overview of agile practices, ideally on a synthesized level of detail. We therefore decided to use the Integrated List of Agile Practices~\cite{Neumann.2022}. The list was created based on a thorough review of the results from a systematic literature review considering the years since 2010 and thus, recent trends in the field. Furthermore, the list comprised agile practices on a similar level of detail as a synthesizing process was applied to harmonize agile practices from a different level of detail. In total, the list consists of 38 agile practices clustered in five categories and provides a profund basis for us in analyzing the relations between agile practices. The integrated list of agile practices is presented in~\cite{Neumann.2022}. Table~\ref{Table:SListoAP_clean} provides an overview of the integrated list of agile practices.

\begin{table}[ht]
\centering
\begin{tabular}{l|p{5.5cm}}
\hline
\textbf{Category} & \textbf{Agile practice} \\
\hline
\multirow{12}{*}{Technical}
 & Agile Testing \\
 & Code review \\
 & Coding standards \\
 & Collective code ownership \\
 & Continuous integration \\
 & DevOps \\
 & Prototyping and Spike Solutions \\
 & Refactoring \\
 & Simple design \\
 & Small and frequent releases \\
 & Software configuration management \\
 & Zero technical depts \\
\hline
\multirow{12}{*}{Collaboration}
  & Agile estimation \\
  & Customer integration \\
  & Co-located team \\
  & Communication \\
  & Daily Standup Meetings \\
  & Pair programming \\
  & Planning Game \\
  & Release Planning \\
  & Retrospective / Learning Loop \\
  & Review Meeting \\
  & Scrum of Scrums \\
\hline
\multirow{4}{*}{Process}
  & Iteration based process \\
  & Limit WIP \\
  & Tracking progress \\
\hline
\multirow{7}{*}{Requirements}
  & Behaviour Driven Development \\
  & Definition of done \\
  & Definition of Ready \\
  & Documentation \\
  & Metaphor / Vision \\
  & User Stories \\
  & Using and maintaining a backlog \\
\hline
\multirow{5}{*}{Organizational}
  & Empowered and self-organizing team \\
  & Energized work \\
  & Knowledge sharing \\
  & Office structure \\
  & Time Boxing \\
\hline
\end{tabular}
\caption{The integrated list of agile practices (without references)}
\label{Table:SListoAP_clean}
\end{table}

In a first step, we created a template for the documentation of the agile practices relations. The template considered the following scheme:

\begin{itemize}
    \item \textit{Description (with source/s):} A explanation of the agile practice including it's primary sources (e.g., a guideline from an agile method).
    \item \textit{Relation reasoning (Coherence with other agile practices):} Based on the description from the literature used in the field above, we argue here the identified relation type(s).
    \item \textit{Adaption classification:} Assessment of whether the agile practice should be used individually or in combination with other practices. When assessing that the practice can be used individually, there must be no requirement relation (see Table~\ref{tab:relationTypes}).
    \item \textit{Empirical validation:} Prepared for a summary from a survey conducted to verify the coherences that we identified on a theoretical basis.
\end{itemize}

We provide below with \textit{AP04 - Collective Code Ownership} a specific example for an agile practice:

\begin{itemize}
    \item \textit{Description (with source/s):} All team members are responsible for all code. Any developer can change any part of the code at any time~\cite{Beck.2000}.
    \item \textit{Relation reasoning (Coherence with other agile practices):} Requires shared coding standards and is supported by practices like version control and communication~\cite{Beck.2000}.
    \item \textit{Adaption classification:} Less effective without common conventions and collaboration structures.
    \item \textit{Empirical validation:} Not done yet.
\end{itemize}

To document the agile practices considered for the development of our model, we conducted a literature review to identify additional studies and references. This process allowed us to obtain concise descriptions and examine the relationships between different agile practices. Our primary focus was on the key studies identified in the tertiary study by Neumann~\cite{Neumann.2022}, as well as the guidelines from established agile frameworks, including Scrum, Kanban, XP, and the Agile Alliance Glossary~\cite{AgileAlliance.2021}.

Subsequently, we analyzed the 38 agile practices from~\cite{Neumann.2022} individually to identify potential borderline cases that might not be suitable for inclusion in our model. As a result, the practices \textit{Software Configuration Management}, \textit{Communication}, \textit{Scrum of Scrums}, and \textit{Energized Work} were excluded from the final set of model-relevant agile practices, warranting further examination. We argue our decision as follows:

\begin{itemize}
    \item \textit{Software Configuration Management} is not considered as an agile practice, as it does not directly serve any of the twelve principles of the agile manifesto.
    \item On the one hand, \textit{Communication} is not seen as a practice in the sense of how something is to be done, and on the other hand, it is a (cultural) factor that is an integrated part of several agile practices, where communication is of importance.
    \item \textit{Scrum of Scrums} is a scaled agile framework rather than a practice.
    \item \textit{Energized Work} defined for XP~\cite{Beck.2000} can be understood as a kind of a hyperonym for a set of agile practices. Due to the abstract description, a large number of practices can be summarized under this generic term.
\end{itemize}

Due to space limitations, the final set of selected agile practices for our model creation is available at Zenodo~\cite{Hansper.2024-1}.

\begin{figure*}[ht]
    \centering
    \includegraphics[scale=0.15]
    {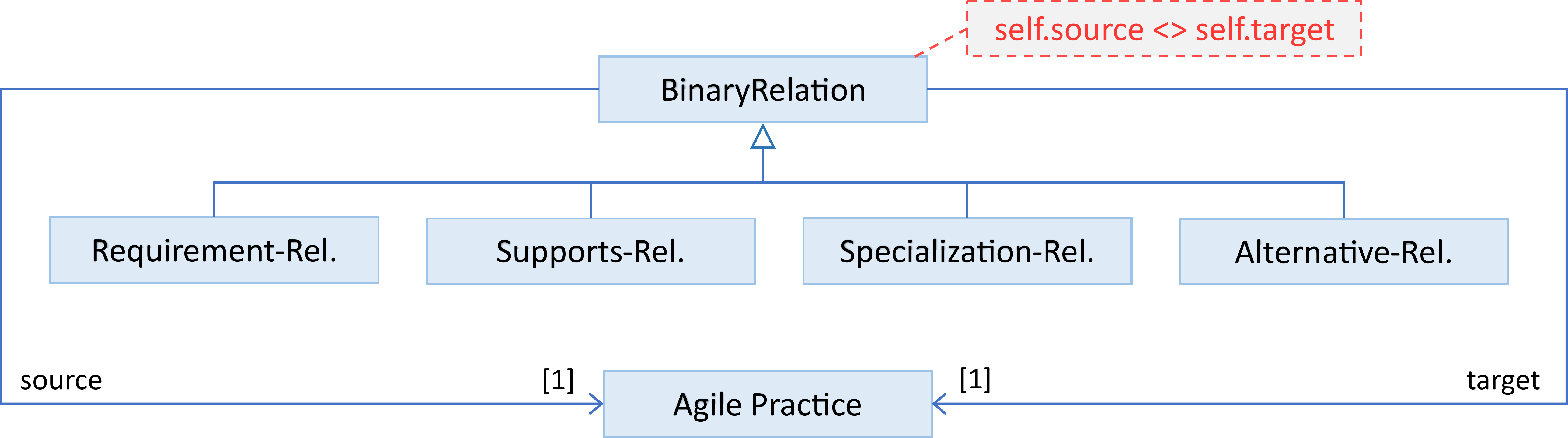}
    \caption{\textit{Agile Map} - meta-model}
    \label{fig:agileMap_metamodel}
\end{figure*}

\section{Relations between Agile Practices}
\label{Sec5:RelationsbetweenAP}
In the next step, we analyzed the agile practices to identify relationships between them. Here, we defined source and target, as we started the analysis based on the association between two agile practices. Initially, the main focus was on identifying so-called requires-relations, which we define as an association with a mandatory character: The effective application of an agile practice requires the application of the associated one.   As time went on, it became clear that other dependencies and thus, relation types also exist. Finally, we identified four types of relationship, which are introduced in Table~\ref{tab:relationTypes}.

\begin{table}[ht]
    \caption{relation types}
    \vspace{-14pt}
    \textcolor{white}{\caption{}}
    \label{tab:relationTypes}
    \centering
    \setlength\extrarowheight{4pt}
    \begin{tabular}{|p{0.11\textwidth}|p{0.32\textwidth}|} 
        \hline
        \rowcolor{tablehead}
        \textbf{graphical}  
        \textbf{representation}
        &
        \vspace{0pt}
        \textbf{description} \\
        \hline
        \begin{wrapfigure}{c}{2.1cm}
            \vspace{-21pt}
            \centering
            \includegraphics[width=2.2cm]
            {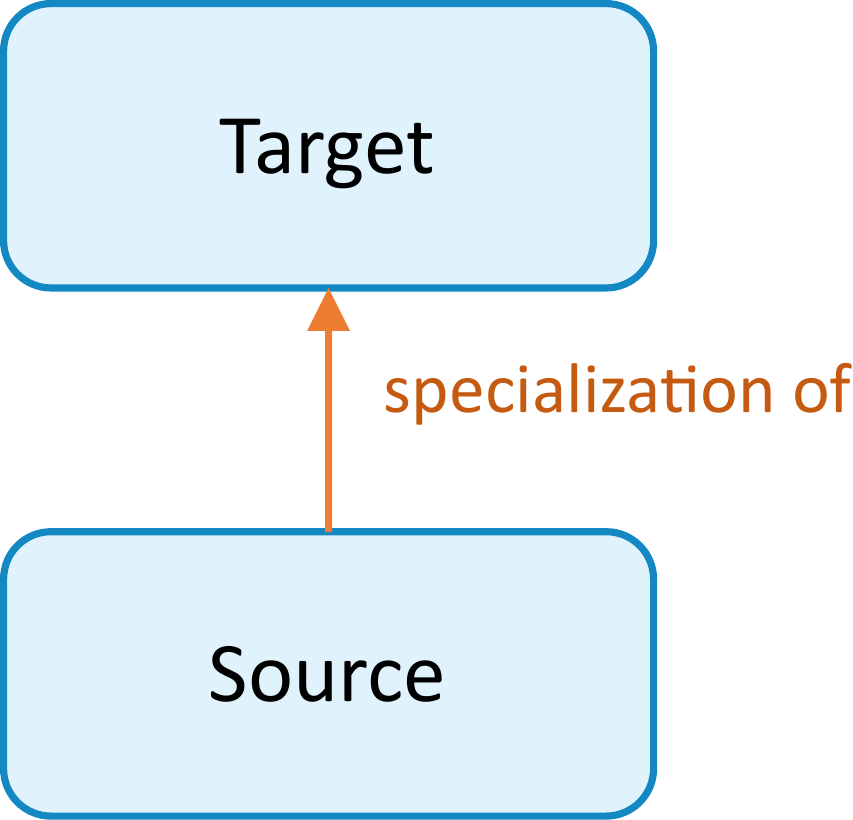}
            \label{fig:Relation_SpecOf}
        \end{wrapfigure}
        &
        The specialization relation describes a subset relationship. The source element therefore represents a special form of the target element.
        \vspace{40pt} \\
        \hline
        \begin{wrapfigure}{c}{2.3cm}
            \vspace{-20pt}
            \centering
            \includegraphics[width=1.7cm]
            {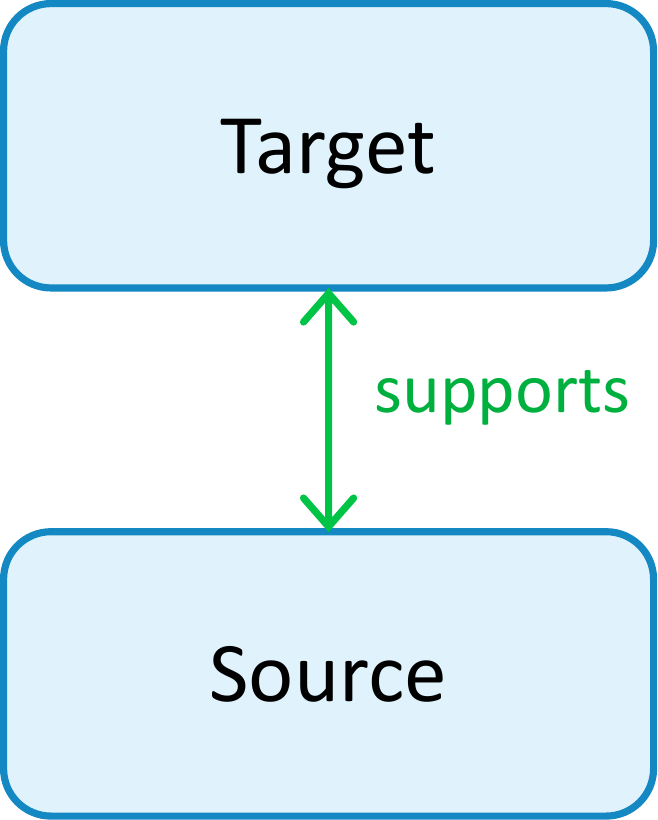}
            \label{fig:Relation_Supp}
        \end{wrapfigure}
        &
        The support relation states that the source element has a supporting effect on the target element. It is used when the target element offers greater added value if the source element is also used.  The relationship can be both unidirectional and bidirectional.
        \vspace{14pt} \\
        \hline
        \begin{wrapfigure}{c}{2.3cm}
            \vspace{-20pt}
            \centering
            \includegraphics[width=1.7cm]
            {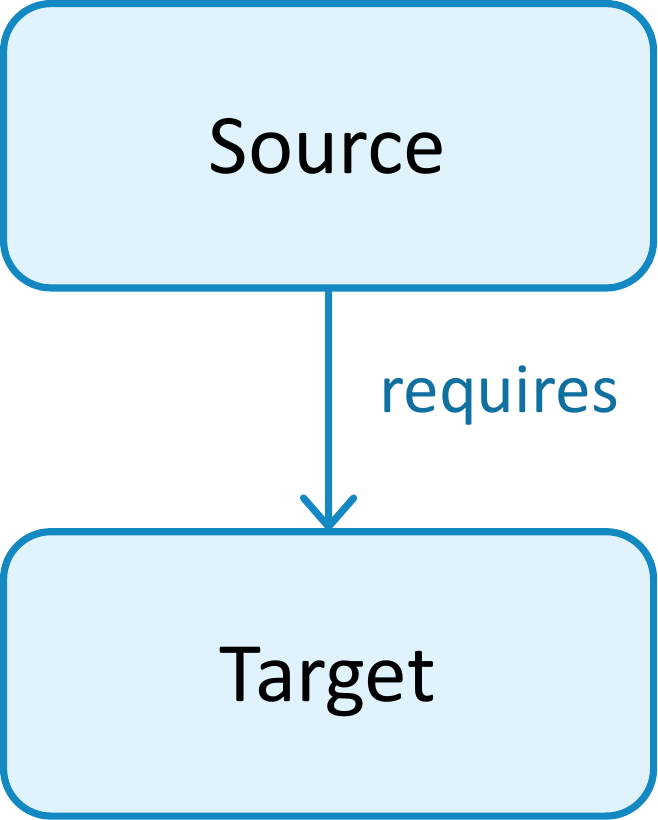}
            \label{fig:Relation_Req}
        \end{wrapfigure}
        &
        The requies relation states that the target element must also be used for the effective implementation of the source element. The mandatory character is also given in this context if the source element can be used without the target element, but this contradicts the basic idea of an agile way of working. The relationship is always unidirectional.
        \vspace{-4pt} \\
        \hline
        \begin{wrapfigure}{c}{2.3cm}
            \vspace{-20pt}
            \centering
            \includegraphics[width=1.7cm]
            {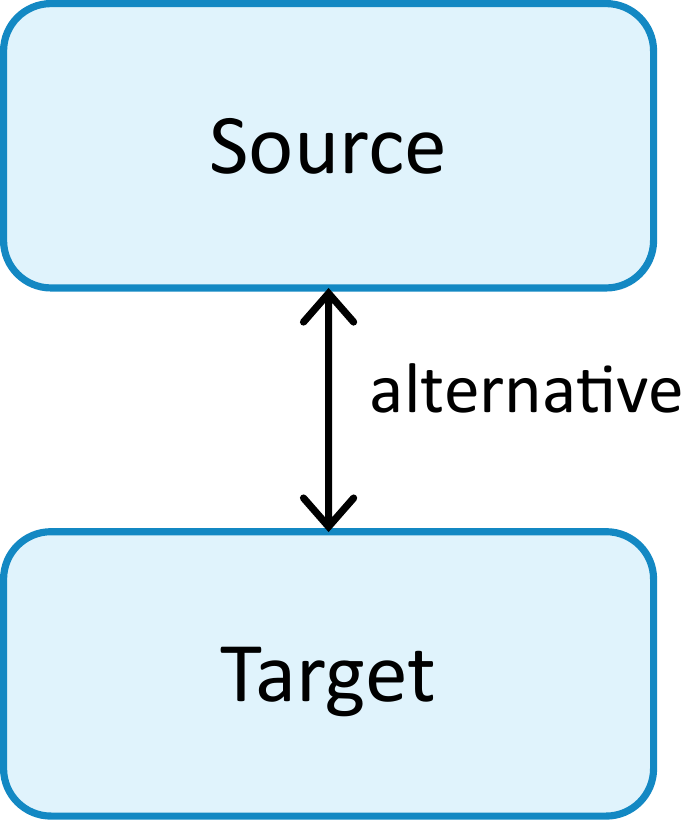}
            \label{fig:Relation_Alt}
        \end{wrapfigure}
        &
        The alternative relation states that the source and target elements are practices that are similar in their basic features and can therefore be used as alternatives to each other. 
        
        The relationship is always bidirectional.
        \vspace{20pt} \\
        \hline
    \end{tabular}
    \vspace{10pt}
    \label{tab:relationTypes}
\end{table}


For its graphical representation, we decided to use the Unified Modeling Language (UML) class diagram as it provides the opportunity to represent all the relation types we defined (see Table~\ref{tab:relationTypes}). Furthermore, it is common to use UML artifacts and diagrams  for the visualization of systematic description of interrelationships and structures~\cite{Diebold.2015a,Neumann.2023}.



\section{The Agile Map}
\label{Sec4:Results}
In this section, we present the core contribution of this study: The Agile Map; a model describing the relationships between agile practices in a systematic manner. 

Before we introduce the Agile Map based on an excerpt, we explain it's systematic structure in form of a meta-model (see Figure~\ref{fig:agileMap_metamodel}). The four relation types are classified as specialized BinaryRelation. It consists of \textit{Binary Relations} that correspond to one of the relation types according to \autoref{tab:relationTypes}. Each relation has exactly one \textit{Agile Practice} as source-element and one \textit{Agile Practice} as target-element. 

If two relations $r_1 \ \land \ r_2$ of the same type exist in such a way that: $source(r_1) = target(r_2) \ \land \ source(r_2) = target(r_1)$, these are combined in the agile map to form a bidirectional relation. The constraint (highlighted in red in Figure~\ref{fig:agileMap_metamodel}) represents the rule, that the source-element must not be the same as the target-element for all relations exist in the Agile Map.

It is important to mention, that the defined coherences between agile practices are on a hypothetical basis. Nevertheless, the coherences were identified and defined in a systematic manner based on a thorough analysis of the literature. Thus, we refer for each agile practice and their coherences to specific literature sources in our repository in~\cite{Hansper.2024}. 
\begin{wrapfigure}{r}{3.8cm}
    \vspace{-10pt}
    \centering
    \includegraphics[height=4.5cm]
    {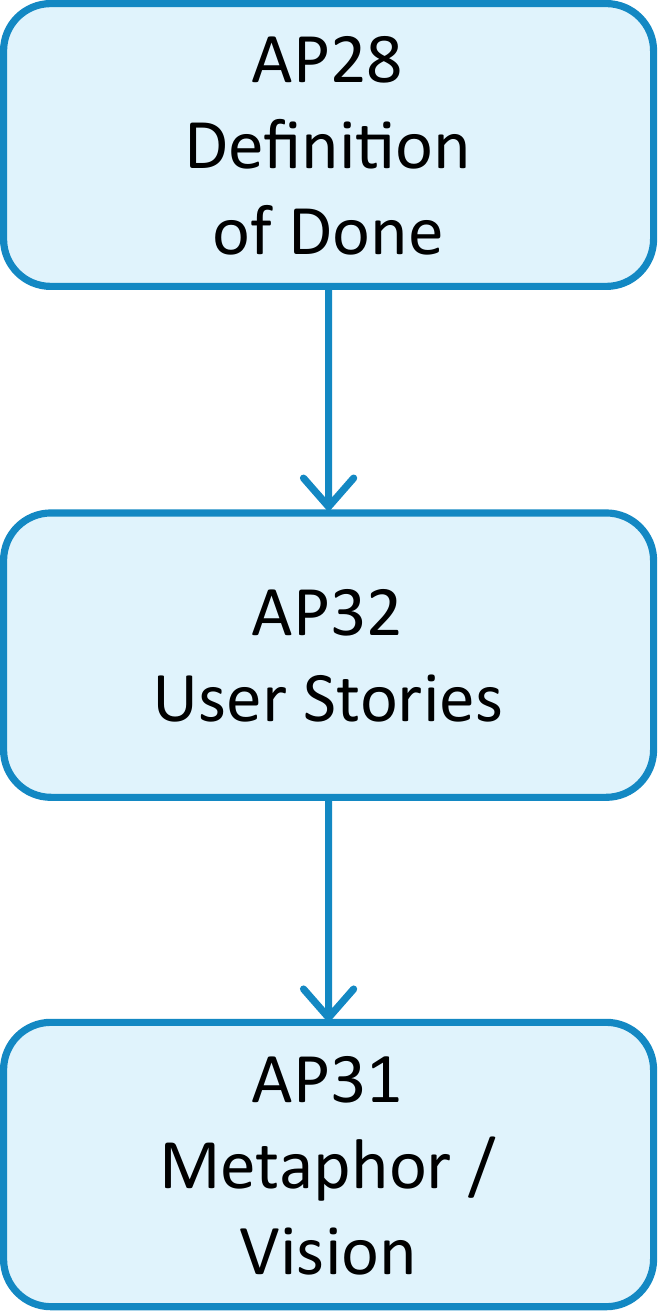}
    \caption{Excerpt of the Agile Map}
    \label{fig:agileMap_excerpt}
\end{wrapfigure}

Even if we are not able to show the full Agile Map due to space limitations, we present an excerpt in Figure~\ref{fig:agileMap_excerpt} to provide an example. However, the full Agile Map is available at Zenodo~\cite{Hansper.2024}.

A \textit{Definition of Done} defines a set of criteria, which represent when task has been (or can understand as) completed. As the specific tasks are derived from \textit{User Story}, this is a mandatory prerequisite for describing the Defintion of Done.

\textit{User Stories} follow the well-known scheme: A user wants a feature to reach an objective. A user can be a dedicated user role or user group. A feature (or function) represents a specific requirement and finally, the the objective of a story is typically refined by economic or technical values. Since the specification of the objective first requires a vision of what should be achieved, the \textit{User Story} requires a relation to the\textit{ Metaphor/Vision}. The references for the literature based coherences among these three agile practices can be found in our repository~\cite{Hansper.2024}.

We conclude: If the source element of a requirement relation is used, all (transitive) target elements must also be applied. The Agile Map consists of 37 agile practices of which 4 agile practices are described as no specific agile practices. In total, we defined 47 theoretical relations between the agile practices. Interestingly, most of the identified relationships (20) are from the type requires, which indicates a strong coherence between them. 

\section{Conclusion \& Future Work}
\label{Sec6:Conclusion}
In this paper we present the Agile Map, a theoretical model describing relations between agile practices. In more detail, we argue the motivation for its creation and our approach to conceptualize the coherences between agile practices.

The Agile Map comprises 37 agile practices, which systematically extracted from the literature using the results from a literature review and a systematic structure to conceptualize their relations to one another. Here, we defined four relation types based on a thorough analysis of the agile practices. To do so, we developed a scheme template documenting the identified relations and the argumentation based on existing literature. In total, the Agile Map defines 47 theoretical relations between agile practices, while the requires-relation is the most used for 20 relations. 

However, even if created the Agile Map based on a systematic approach, we are aware of its limitations. In its current version, the relations are described in a hypothetical manner. Thus, a validation of the relations using empirical data is necessary. We therefore already preparing a questionnaire survey aiming to validate the identified relations and to be able to optimize the Agile Map. 

\bibliographystyle{IEEEtran}
\bibliography{references}

\end{document}